# Revisiting the Analytical Solution of Spin-Orbit Torque Switched Nanoscale Perpendicular Ferromagnet

Xue Zhang[1,2,3], Zhengde Xu[1,2,3], and Zhifeng Zhu[1,4†]

[1]School of Information Science and Technology, ShanghaiTech University, Shanghai, China 201210

[2]Shanghai Institute of Microsystem and Information Technology, Chinese Academy of Sciences, Shanghai, China, 200050

[3]University of Chinese Academy of Sciences, Beijing, China, 100049

[4]Shanghai Engineering Research Center of Energy Efficient and Custom AI SIC, Shanghai, China 201210

## Abstract

The scaling of magnetic memory into nanometer size calls for a theoretical model to accurately predict the switching current. Previous models show large discrepancy with experiments in studying the spin-orbit torque switching of perpendicular magnet. In this work, we find that the trajectory of magnetization shows a smooth transition during the switching. This contradicts the key assumption in previous models that magnetization needs to align to the spin polarization ($\sigma$) before the switching occurs. We demonstrate that aligning magnetization to $\sigma$ requires a very large current, resulting in the unsatisfactory fitting between the previous models and experiments. In contrast, the smooth transition permits a lower switching current that is comparable to experiments. Guided by this refined physical picture, we pinpoint the reversal of precession chirality as a pivotal factor for achieving deterministic switching. This insight leads to the formulation of a new analytical model that demonstrates remarkable agreement with experimental data. Our work resolves an important issue confronting the experiment and theory. We provide a clear physical picture of the current-induced magnetization switching, which is invaluable in the development of spin-orbit torque magnetic random-access memory.

## Introduction

In recent years, significant progress has been achieved in the manipulation of magnetization through magnetic field [1-5], spin-transfer torque (STT) [6-16], and spin-orbit torque (SOT) [17-37]. However, the field switched device is not conducive to integration. The lifetime of the device driven by STT is greatly diminished because the current passes directly through the MTJ. As a result, researchers are turning to SOT-induced switching, which offers many advantages such as a lower writing current and faster operation speeds [22, 23, 38-40]. Depending on the different arrangement of the spin polarization ($\sigma$) and the magnetization, the SOT switching can be classified into three types, i.e., type *x*, *y* and *z* [38]. In type *y*, the magnetization can be directly switched between +**y** and −**y** since $\sigma$ is collinear to the magnetization [41]. However, the threshold current density ($J_{th}$) is large since the demagnetizing field impedes the switching. In addition, type *x* and *y* require an elliptical sample to stabilize the magnetization in the thin film plane, which presents challenges in the device fabrication when a small sample to sample variation is required. In contrast, the demagnetizing field assists the switching in type *z*, and the device can have circular shape since the in-plane components along the **x** and **y** axes are on an equal footing.

Since $\sigma$ and the equilibrium magnetization are noncollinear in type *z*, the deterministic switching requires both the SOT and the assisted magnetic field along the current direction, and one cannot simply balance the SOT and damping-like torque to obtain $J_{th}$. In previous studies, an analytical equation, $J_{th} = \frac{2e}{\hbar} \frac{M_s t_F}{\theta_{SH}} \left( \frac{H_K^{eff}}{2} - \frac{H_{ext}}{\sqrt{2}} \right)$, has been proposed based on the macrospin model and assumes that $m_y$ is first aligned to $\sigma$ [42]. In this equation, $H_K^{eff} = H_k - H_d$ [43] is the effective anisotropy field. Based on the same principle, modified equations are proposed to include the effect of field-like torque (FLT), thermal fluctuation and second-order perpendicular magnetic anisotropy [44-46]. However, $J_{th}$ predicted by these equations is much higher than the experimental values [47-52]. This discrepancy is often ascribed to the micromagnetic nature of the experimental results. The experimental results show that $J_{th}$ of micron-scale devices ranges from 2 to 9 × 10$^{11}$ A/m$^2$, which is lower than that of nanoscale devices (2 to 6 ×

$10^{12}$ A/m$^2$). This reduction in $J_{th}$ in larger samples is attributed to nonuniform switching caused by the domain nucleation. However, as the device size shrinks, it is imperative to evaluate the accuracy of this analytical model in the nanoscale device. To this end, delicate experiments have been carried out [53, 54]. However, they concluded that the fitting using these analytical models leads to unrealistic parameters such as a very large $\theta_{SH}$.

In this work, using the macrospin model, we present numerical results of SOT switching of perpendicular magnet. The trajectory of magnetization shows a smooth transition from the initial state to the opposite hemisphere, followed by the precessional damping to the other stable state. It is worth noticing that magnetization does not align to **σ** during this process, which violates the assumption used in the previous analytical models. The smooth transition indicates a smaller switching current in the numerical and analytical models, which we find are comparable to the experimental values. Based on this observation, we present a theoretical analysis for the switching, and we identify the alteration of precession chirality as the key factor for the magnetization switching. We show that our numerical and analytical results agree perfectly with the experimental results for the nanoscale devices. In addition, we show our results are insensitive to both small and large damping constant.

## Methodology

The device studied in this work is illustrated in Fig. 1(a). The FM layer has perpendicular magnetization that is stabilized by the crystalline anisotropy field (**H**$_K$) along **z** direction and demagnetizing field (**H**$_d$) which is calculated based on the sample geometry (50 nm×50 nm×1.2 nm). Passing the charge current (**J**$_c$) along **x** direction through the heavy-metal (HM) layer produces a spin current (**J**$_s$) that flows into the ferromagnetic (FM) layer with a resultant SOT. The magnetization dynamics is described by the macrospin model using the Landau-Lifshitz-Gilbert-Slonczewski (LLGS) equation:

$$\frac{\partial \mathbf{m}}{\partial t} = -\gamma \mathbf{m} \times (\mathbf{H}_{eff} + \mathbf{H}_{ext}) + \alpha \mathbf{m} \times \frac{\partial \mathbf{m}}{\partial t} + B_{D,SOT} \mathbf{m} \times (\mathbf{m} \times \boldsymbol{\sigma}),$$

which is numerically integrated through the fourth-order Runge-Kutta methods (RKMs). $\gamma = 1.76 \times 10^{11}$ s$^{-1}$T$^{-1}$ is the gyromagnetic ratio, and $\mathbf{H}_{\text{eff}} = \mathbf{H}_K + \mathbf{H}_d = 0.5$ T is the effective magnetic field. The damping constant $\alpha = 0.02$ [53]. $B_{\text{D, SOT}} = \frac{\hbar \theta_{\text{SH}} J_c}{2eM_s t_F}$ represents the strength of the damping-like SOT. The spin-Hall angle $\theta_{\text{SH}} = 0.03$ [53] and the saturation magnetization $M_s = 1.3$ T [53]. $t_F$ is the thickness of the FM layer. The direction of $\boldsymbol{\sigma}$ can be determined by $\mathbf{J}_s = \theta_{\text{SH}} \boldsymbol{\sigma} \times \mathbf{J}_c$, i.e., $\boldsymbol{\sigma} = -\mathbf{y}$ when $\mathbf{J}_c$ is along $\mathbf{x}$ direction.

**Result and Discussion**

The SOT switching of perpendicular FM has a key feature that both SOT and $\mathbf{H}_{\text{ext}}$ are indispensable. We verify this by separately applying $\mathbf{J}_c$ and $\mathbf{H}_{\text{ext}}$. As shown in Fig. 1(b), when only $\mathbf{H}_{\text{ext}}$ is applied, the magnetization remains at the initial state. As shown in Fig. 1(c), in the absence of $\mathbf{H}_{\text{ext}}$, the magnetization remains at the initial state under small $\mathbf{J}_c$, and the SOT pulls it to $\boldsymbol{\sigma}$ when $\mathbf{J}_c$ is very large at $4.5 \times 10^{12}$ A/m$^2$. In contrast, when a small $\mathbf{H}_{\text{ext}} = 20$ mT is applied, the magnetization can be easily switched under a small $\mathbf{J}_c = 2.604 \times 10^{12}$ A/m$^2$ [see Fig. 1(d)]. In addition, the switching polarity reverses when $\mathbf{H}_{\text{ext}}$ is reversed. These results show that our model correctly captures the key features of the SOT induced switching. It is worth noting that the experimental switching current density is $2.76 \times 10^{12}$ A/m$^2$ [53, 54], which is almost the same as our result. In contrast, the previous model predicts $J_{\text{th}} = \frac{2e}{\hbar} \frac{M_s t_F}{\theta_{\text{SH}}} \left( \frac{H_K^{\text{eff}}}{2} - \frac{H_x}{\sqrt{2}} \right) = 76 \times 10^{12}$ A/m$^2$ [42], in which $H_K^{\text{eff}} = 0.5$ T is used. This value is 28 times larger than experimental results. Since samples with diameter $D = 50$nm were used in the experiment, one can assume that it is uniformly switched. Therefore, our numerical results agree excellently with the experiment, indicating that the macrospin model is indeed suitable to describe the magnetization switching in the nanoscale magnet. The remaining problem is to resolve the disagreement between the analytical model and the experimental results.

Recall that the analytical model in Ref. 42 requires the magnetization to first align to $\boldsymbol{\sigma}$ before switching to the other state, as illustrated in Fig. 2(a). It has been shown in

Fig. 1(c) that pulling the magnetization to **σ** requires a large $J_c$, i.e., $4.5 \times 10^{12}$ A/m². This reveals the reason that the previous analytical model always predicts a much larger $J_{th}$. In contrast, as shown in Fig. 2(b), the switching trajectory from numerical simulation shows a smooth magnetization transition from up to down [38], in which the magnetization does not have an intermediate state aligning to **σ**. The smooth transition explains the smaller $J_{th}$ obtained in the numerical simulation that is identical to the experimental results. Comparing these two switching pictures, one can immediately realize the key to resolve the disagreement between the experiment and the analytical model is to develop a model based on the smooth switching picture.

To determine the critical switching condition, we analyzed the magnetization trajectory before and after $J_{th}$ [marked as a square and star in Fig. 1(d)]. As shown in Figs. 2(c) and 2(d), when $J_c$ is close to $J_{th}$, the magnetization precesses without triggering the switching. It is noticed that the precession direction remains in the counterclockwise direction. When $J_c = J_{th}$, the switching happens, and the trajectories are shown in Figs. 1(e) and 2(b), in which we define a point marked by a star. Before this point, the magnetization precesses in the counterclockwise direction, the same as the previous non-switching case. However, after the magnetization passes this point, the precession direction reverses. We have varied both the polarity and magnitude of **J**$_c$ and **H**$_{ext}$, where we find the successful switching is always accompanied by the reversal of precession chirality. Therefore, we identify it as one of the critical conditions for deterministic switching. In addition, we plot the evolution of magnetization in the first 5 ns in Fig. 3(a). It is clearly seen that in the plateau region (marked by brown) close to the critical point, $m_x$, $m_y$ and $m_z$ are stable, and $m_y$ is almost zero (i.e., the azimuthal angle $\varphi = 0$). We have verified this by simulating the magnetization switching under different **H**$_{ext}$. As shown in Fig. 3(b), the polar angle ($\theta$) is linearly increased, whereas $\varphi$ is always zero. We use this as the other critical condition of the deterministic switching.

To derive the analytical expression of $J_{th}$ based on these observations, we separate **H**$_{ext}$ from the effective field and convert the LLG equation into the Landau-Lifshitz (LL) form:

$$\frac{\partial \mathbf{m}}{\partial t}(1+\alpha^2) = -\gamma \mathbf{m} \times \mathbf{H}_{ext} - \gamma\alpha \mathbf{m} \times (\mathbf{m} \times \mathbf{H}_{ext}) - \gamma \mathbf{m} \times \mathbf{H}_{eff'} - \gamma\alpha \mathbf{m} \times (\mathbf{m} \times \mathbf{H}_{eff'})$$
$$+ \alpha\gamma B_{D,SOT} \mathbf{m} \times \boldsymbol{\sigma} - \gamma B_{D,SOT} \mathbf{m} \times (\mathbf{m} \times \boldsymbol{\sigma}),$$

where $\mathbf{H}_{eff'}$ includes only the anisotropy and demagnetizing field. As a result, there are six torques in this system which are illustrated in Fig. 3(c). We start in the state with $\mathbf{m}$ along the +**z**-direction, where torque 3 ($-\gamma\mathbf{m}\times\mathbf{H}_{eff'}$) and torque 4 ($-\gamma\alpha\mathbf{m}\times(\mathbf{m}\times\mathbf{H}_{eff'})$) are zero. The rest of the torques pull $\mathbf{m}$ into x(-y)z vector space as depicted in the inset with gray arrows. Subsequently, $\mathbf{m}$ gradually moves closer to the x-z plane until the chirality reverses. The related torque analysis in this critical point in the x-z plane is depicted in Fig. 3(d). It is clear to see that the rotation direction is determined by torques 1, 3, and 6 while torques 2, 4, and 5 compete with each other to switch $\mathbf{m}$ across the x-y plane. We then express $\mathbf{m} = [\sin\theta\cos\varphi, \sin\theta\sin\varphi, \cos\theta]$, and $\mathbf{H}_{eff'} = [-H_{d,x}m_x, -H_{d,y}m_y, (H_k-H_{d,z})m_z]$ where $H_d$ is the strength of the demagnetizing field. The specific expression of the rotation torques are listed in Table 1. To fulfill the condition that the precession chirality is reversed, one needs the total torque from 1, 3, and 6 to be zero. With the additional requirement that $\varphi = 0$, we arrive at an analytical expression of $J_{th}$ as:

$$J_{th} = \frac{\cos\theta(\sin\theta(H_k - H_{d,z} + H_{d,x}) - H_{ext})2t_{FM}M_s e}{\theta_{SH}\hbar} \quad (4).$$

Table 1: Expressions of torque 1, 3, and 6 in different directions

|   | Torque 1 | Torque 3 | Torque 6 |
|---|---|---|---|
| x | 0 | $m_z H_{eff,y} - m_y H_{eff,z}$ | $B_{D,SOT}(m_x m_y)$ |
| y | $-m_z H_{ext}$ | $m_x H_{eff,z} - m_z H_{eff,x}$ | $B_{D,SOT}(-m_z^2 - m_x^2)$ |
| z | $m_y H_{ext}$ | $m_y H_{eff,x} - m_x H_{eff,y}$ | $B_{D,SOT}(m_z m_y)$ |

We then compare our results with experimental data in devices of various sizes. As shown in Fig. 4(a), it is clearly seen that our analytical and macrospin simulation results agree well with the experimental data, whereas the previous analytical model predicts a $J_{th}$ that is 28 times larger. In addition, this equation is damping-independent which arises from the torque balance. As shown in Fig. 4(b), the simulation results also reveal that the damping constant has negligible effect on $J_{th}$. A remarkable result is that we show $J_{th}$ remains the same for any value of $α$, whereas difference with previous study [42] concludes that it only appears in the system with large $α$. Furthermore, they excluded the influence of damping based on the observations from the numerical results, whereas we have included the damping in our derivation and find it does not affect our results. Additionally, when the device diameter ($D$) is increased, a decrease in $J_{th}$ is captured by our model, which is correlated with $θ$ at the critical point. We also studied the impact of field-like spin-orbit torque (FLT) on the system. $β$ is defined as the ratio between the FLT and DLT, which have opposite signs [55]. The resulting $J_{th}$ is summarized in Fig. 4(b) with $H_{ext}$ = 20 mT and $D$ = 50 nm. We found that when $β$ is increased, the corresponding $J_{th}$ first increases and then decreases. The similar variation has been reported in other works [56, 57]. To understand the reason for this change, we plotted the switching process in the x-y plane for systems with $β$ = –0.2 and $β$ = –3 in Fig. 4(c). Compared to Fig. 2(d), the chirality reversal point still exists, indicating that we can still obtain $J_{th}$ using the same method. The main difference is that the chirality reversal no longer occurs in the x-z plane; instead, it happens in the x(-y)z vector space. The corresponding torque analysis is shown in Fig. 4(d). We found that FLT introduces two torques, torque 7 and torque 8. Since $β$ is negative, torque 7 is always opposite to torque 5 which prevents **m** from reaching x-z plane. As a result, the reversal of chirality occurs in the x(-y)z vector space as shown in Fig. 4(c). The direction of rotation in this space is determined by the competition among torques 1, 2, 3, 6, and 8. When $β$ is small, torque 8 is negligible. The rotation direction is determined by torques 1 and 6 compete with torques 2 and 3. Compared to the sample when $β$ = 0 shown in Fig. 3(d), torque 2 appears and acts as an additional counterclockwise rotational factor in the x(-y)z space. To balance this counterclockwise influence, torques 1 and 6 must be increased. Since $H_{ext}$ is fixed, the only feasible way is by increasing $J_c$. As β increases, torque 7 also increases, preventing **m** from reaching the x-z and x-y planes. Consequently, the angle between **m** and $H_{ext}$ increases, leading to an increase in torques 1 and 2, and resulting in a higher $J_{th}$. However, when $\beta$ is further increased, torque 8 becomes significant and assists torque 6 in the switching process, making the switching much easier. However, since the critical point is located in the x(-y)z vector space, one cannot use $\varphi = 0$ to obtain a simple analytical expression. The final formula becomes overly complex. Nonetheless, it is worth noting that the change in $J_{th}$ is relatively small when a finite $\beta$ is included, e.g., $(J_{th,\beta=-2}-J_{th,\beta=0})/J_{th,\beta=0} = 20\%$. Moreover, we also consider the effect of the thermal fluctuation, which is included as a random field [58] $\mu_0 H_{thermal} = N(0,\mu)\hat{x} + N(0,\mu)\hat{y} + N(0,\mu)\hat{z}$ where $N(0,\mu)$ represents the normal distribution. $\mu = \sqrt{2k_B T\alpha/(V_{FL}M_s(1+\alpha^2)\Delta t)}$ is the standard deviations. $T = 300$K, $V_{FL}$ is the volume of the free layer and $\Delta t$ is the duration of the thermal fluctuations. We find $J_{th}$ is slightly reduced from $2.604\times10^{12}$ A/m² to $2.49\times10^{12}$ A/m². This allows us to use a smaller $H_{eff}$, permitting a better agreement with experiments.

## Conclusion

In this work, we revisit the spin-orbit torque switching of perpendicular magnet in nanoscale. We emphasize that the spin-orbit torque switching of perpendicular magnet cannot be visualized as a two-step process, i.e., the magnetization first aligns to **σ** and then switches. In contrast, the spin-orbit torque and external magnetic field act cooperatively, producing a smooth magnetization transition. The refined physical picture guarantees a smaller switching current which is comparable to the experimental results. We further identify the chirality reversal of precession as a key condition for the deterministic switching, based on which we derived a new analytical expression for the switching current that shows remarkable agreement with experiments.

## Acknowledgment

We acknowledge the support from the National Key R&D Program of China

(Grant No. 2022YFB4401700), National Natural Science Foundation of China (Grants Nos. 12104301 and 62074099). The simulation conducted in this work is supported by SIST Computing Platform at ShanghaiTech University.

## References

1 H. Ohno, a. D. Chiba, a. F. Matsukura, T. Omiya, E. Abe, T. Dietl, Y. Ohno, K. Ohtani, Electric-field control of ferromagnetism, Nature. **408**, 944 (2000).
2 S. Kanai, M. Yamanouchi, S. Ikeda, Y. Nakatani, F. Matsukura, H. Ohno, Electric field-induced magnetization reversal in a perpendicular-anisotropy CoFeB-MgO magnetic tunnel junction, Appl. Phys. Lett. **101**, 122403 (2012).
3 W.-G. Wang, M. Li, S. Hageman, C. Chien, Electric-field-assisted switching in magnetic tunnel junctions, Nat. Mater. **11**, 64 (2012).
4 Y. Shiota, T. Nozaki, F. Bonell, S. Murakami, T. Shinjo, Y. Suzuki, Induction of coherent magnetization switching in a few atomic layers of FeCo using voltage pulses, Nat. Mater. **11**, 39 (2012).
5 A. Haldar, A. O. Adeyeye, Microwave assisted gating of spin wave propagation, Appl. Phys. Lett. **116**, 162403 (2020).
6 L. Berger, Emission of spin waves by a magnetic multilayer traversed by a current, Phys. Rev. B. **54**, 9353 (1996).
7 M. Tsoi, A. Jansen, J. Bass, W.-C. Chiang, M. Seck, V. Tsoi, P. Wyder, Excitation of a magnetic multilayer by an electric current, Phys. Rev. Lett. **80**, 4281 (1998).
8 E. Myers, D. Ralph, J. Katine, R. Louie, R. Buhrman, Current-induced switching of domains in magnetic multilayer devices, Science. **285**, 867 (1999).
9 J. Z. Sun, Spin-current interaction with a monodomain magnetic body: A model study, Phys. Rev. B. **62**, 570 (2000).
10 Z. Yuan, J. Long, Z. Xu, Y. Xin, L. An, J. Ren, X. Zhang, Y. Yang, Z. Zhu, Anomalous impact of thermal fluctuations on spin transfer torque induced ferrimagnetic switching, J. Appl. Phys. **133**, 153903 (2023).
11 X. Zhang, Z. Xu, J. Ren, Y. Qiao, W. Fan, Z. Zhu, Spin-transfer-torque induced spatially nonuniform switching in ferrimagnets, Appl. Phys. Lett. **124**, 012405 (2024).
12 X. Zhang, B. Cai, J. Ren, Z. Yuan, Z. Xu, Y. Yang, G. Liang, Z. Zhu, Spatially nonuniform oscillations in ferrimagnets based on an atomistic model, Phys. Rev. B. **106**, 184419 (2022).
13 J.-G. Zhu, A. Shadman, Resonant Spin-Transfer Torque Magnetoresistive Memory, IEEE Trans. Magn. **55**, 1 (2018).
14 T. Böhnert, Y. Rezaeiyan, M. Claro, L. Benetti, A. Jenkins, H. Farkhani, F. Moradi, R. Ferreira, Weighted spin torque nano-oscillator system for neuromorphic computing, Communications Engineering. **2**, 65 (2023).
15 H. Xie, X. Chen, Q. Zhang, Z. Mu, X. Zhang, B. Yan, Y. Wu, Magnetization switching in polycrystalline Mn3Sn thin film induced by self-generated spin-polarized current, Nat. Commun. **13**, 5744 (2022).
16 D.-F. Shao, Y.-Y. Jiang, J. Ding, S.-H. Zhang, Z.-A. Wang, R.-C. Xiao, G. Gurung, W. Lu, Y. Sun, E. Y. Tsymbal, Néel spin currents in antiferromagnets, Phys. Rev. Lett. **130**, 216702 (2023).

17 A. Brataas, A. D. Kent, H. Ohno, Current-induced torques in magnetic materials, Nat. Mater. **11**, 372 (2012).

18 A. Chernyshov, M. Overby, X. Liu, J. K. Furdyna, Y. Lyanda-Geller, L. P. Rokhinson, Evidence for reversible control of magnetization in a ferromagnetic material by means of spin–orbit magnetic field, Nature Physics. **5**, 656 (2009).

19 C.-F. Pai, L. Liu, Y. Li, H. Tseng, D. Ralph, R. Buhrman, Spin transfer torque devices utilizing the giant spin Hall effect of tungsten, Appl. Phys. Lett. **101**, 122404 (2012).

20 S. Emori, U. Bauer, S.-M. Ahn, E. Martinez, G. S. Beach, Current-driven dynamics of chiral ferromagnetic domain walls, Nat. Mater. **12**, 611 (2013).

21 M. Yamanouchi, L. Chen, J. Kim, M. Hayashi, H. Sato, S. Fukami, S. Ikeda, F. Matsukura, H. Ohno, Three terminal magnetic tunnel junction utilizing the spin Hall effect of iridium-doped copper, Appl. Phys. Lett. **102**, 212408 (2013).

22 I. M. Miron, K. Garello, G. Gaudin, P.-J. Zermatten, M. V. Costache, S. Auffret, S. Bandiera, B. Rodmacq, A. Schuhl, P. Gambardella, Perpendicular switching of a single ferromagnetic layer induced by in-plane current injection, Nature. **476**, 189 (2011).

23 L. Liu, C.-F. Pai, Y. Li, H. Tseng, D. Ralph, R. Buhrman, Spin-torque switching with the giant spin Hall effect of tantalum, Science. **336**, 555 (2012).

24 Z. Xu, X. Zhang, Y. Qiao, G. Liang, S. Shi, Z. Zhu, Deterministic spin-orbit torque switching including the interplay between spin polarization and kagome plane in $Mn_3Sn$, Phys. Rev. B. **109**, 134433 (2024).

25 Z. Xu, J. Ren, Z. Yuan, Y. Xin, X. Zhang, S. Shi, Y. Yang, Z. Zhu, Field-free spin–orbit torque switching of an antiferromagnet with perpendicular Néel vector, J. Appl. Phys. **133**, 153904 (2023).

26 H. Wu, J. Nance, S. A. Razavi, D. Lujan, B. Dai, Y. Liu, H. He, B. Cui, D. Wu, K. Wong, Chiral symmetry breaking for deterministic switching of perpendicular magnetization by spin–orbit torque, Nano letters. **21**, 515 (2020).

27 P. Kumar, A. Naeemi, Benchmarking of spin–orbit torque vs spin-transfer torque devices, Appl. Phys. Lett. **121**, 112406 (2022).

28 H. Vakili, S. Ganguly, G. J. de Coster, M. R. Neupane, A. W. Ghosh, Low Power In-Memory Computation with Reciprocal Ferromagnet/Topological Insulator Heterostructures, ACS nano. **16**, 20222 (2022).

29 K.-J. Kim, S. K. Kim, Y. Hirata, S.-H. Oh, T. Tono, D.-H. Kim, T. Okuno, W. S. Ham, S. Kim, G. Go, Fast domain wall motion in the vicinity of the angular momentum compensation temperature of ferrimagnets, Nat. Mater. **16**, 1187 (2017).

30 N. Roschewsky, C.-H. Lambert, S. Salahuddin, Spin-orbit torque switching of ultralarge-thickness ferrimagnetic GdFeCo, Phys. Rev. B. **96**, 064406 (2017).

31 R. Rahman, S. Bandyopadhyay, A Nonvolatile All-Spin Nonbinary Matrix Multiplier: An Efficient Hardware Accelerator for Machine Learning, IEEE Transactions on Electron Devices. **69**, 7120 (2022).

32 Y. M. Hung, Y. Shiota, S. Yamada, M. Ohta, T. Shibata, T. Sasaki, R. Hisatomi, T. Moriyama, T. Ono, Positive correlation between interlayer exchange coupling and the driving current of domain wall motion in a synthetic antiferromagnet, Appl. Phys. Lett. **119**, 032407 (2021).

33 L. Liao, F. Xue, L. Han, J. Kim, R. Zhang, L. Li, J. Liu, X. Kou, C. Song, F. Pan, Efficient orbital torque in polycrystalline ferromagnetic-metal/Ru/$Al_2O_3$ stacks: Theory and experiment, Phys. Rev. B. **105**, 104434 (2022).

34 Y. Shu, Q. Li, J. Xia, P. Lai, Z. Hou, Y. Zhao, D. Zhang, Y. Zhou, X. Liu, G. Zhao, Realization of the skyrmionic logic gates and diodes in the same racetrack with enhanced and modified edges, Appl. Phys.

Lett. **121**, 042402 (2022).

35 J. Lu, W. Li, J. Liu, Z. Liu, Y. Wang, C. Jiang, J. Du, S. Lu, N. Lei, S. Peng, Voltage-gated spin-orbit torque switching in IrMn-based perpendicular magnetic tunnel junctions, Appl. Phys. Lett. **122**, 012402 (2023).

36 T. Xu, H. Bai, Y. Dong, L. Zhao, H.-A. Zhou, J. Zhang, X.-X. Zhang, W. Jiang, Robust spin torque switching of noncollinear antiferromagnet Mn3Sn, APL Materials. **11**, 071116 (2023).

37 K. Y. Camsari, R. Faria, B. M. Sutton, S. Datta, Stochastic p-bits for invertible logic, Phys. Rev. X. **7**, 031014 (2017).

38 S. Fukami, T. Anekawa, C. Zhang, H. Ohno, A spin–orbit torque switching scheme with collinear magnetic easy axis and current configuration, Nat. Nanotechnol. **11**, 621 (2016).

39 X. Qiu, Z. Shi, W. Fan, S. Zhou, H. Yang, Characterization and manipulation of spin orbit torque in magnetic heterostructures, Advanced Materials. **30**, 1705699 (2018).

40 J. Yoon, S.-W. Lee, J. H. Kwon, J. M. Lee, J. Son, X. Qiu, K.-J. Lee, H. Yang, Anomalous spin-orbit torque switching due to field-like torque–assisted domain wall reflection, Science Advances. **3**, e1603099 (2017).

41 Y.-W. Oh, S.-h. Chris Baek, Y. Kim, H. Y. Lee, K.-D. Lee, C.-G. Yang, E.-S. Park, K.-S. Lee, K.-W. Kim, G. Go, Field-free switching of perpendicular magnetization through spin–orbit torque in antiferromagnet/ferromagnet/oxide structures, Nat. Nanotechnol. **11**, 878 (2016).

42 K.-S. Lee, S.-W. Lee, B.-C. Min, K.-J. Lee, Threshold current for switching of a perpendicular magnetic layer induced by spin Hall effect, Appl. Phys. Lett. **102**, 112410 (2013).

43 H. Sato, E. Enobio, M. Yamanouchi, S. Ikeda, S. Fukami, S. Kanai, F. Matsukura, H. Ohno, Properties of magnetic tunnel junctions with a MgO/CoFeB/Ta/CoFeB/MgO recording structure down to junction diameter of 11 nm, Appl. Phys. Lett. **105**, 062403 (2014).

44 T. Taniguchi, S. Mitani, M. Hayashi, Critical current destabilizing perpendicular magnetization by the spin Hall effect, Phys. Rev. B. **92**, 024428 (2015).

45 S. J. Yun, K.-J. Lee, S. H. Lim, Critical switching current density induced by spin Hall effect in magnetic structures with first-and second-order perpendicular magnetic anisotropy, Sci. Rep. **7**, 15314 (2017).

46 K.-S. Lee, S.-W. Lee, B.-C. Min, K.-J. Lee, Thermally activated switching of perpendicular magnet by spin-orbit spin torque, Appl. Phys. Lett. **104**, 072413 (2014).

47 A. Manchon, J. Železný, I. M. Miron, T. Jungwirth, J. Sinova, A. Thiaville, K. Garello, P. Gambardella, Current-induced spin-orbit torques in ferromagnetic and antiferromagnetic systems, Reviews of Modern Physics. **91**, 035004 (2019).

48 S. Fukami, C. Zhang, S. DuttaGupta, A. Kurenkov, H. Ohno, Magnetization switching by spin–orbit torque in an antiferromagnet–ferromagnet bilayer system, Nat. Mater. **15**, 535 (2016).

49 E. Grimaldi, V. Krizakova, G. Sala, F. Yasin, S. Couet, G. Sankar Kar, K. Garello, P. Gambardella, Single-shot dynamics of spin–orbit torque and spin transfer torque switching in three-terminal magnetic tunnel junctions, Nat. Nanotechnol. **15**, 111 (2020).

50 L. Zhu, Switching of perpendicular magnetization by spin–orbit torque, Advanced Materials. **35**, 2300853 (2023).

51 L. Ren, C. Zhou, X. Song, H. T. Seng, L. Liu, C. Li, T. Zhao, Z. Zheng, J. Ding, Y. P. Feng, Efficient Spin–Orbit Torque Switching in a Perpendicularly Magnetized Heusler Alloy MnPtGe Single Layer, ACS nano. **17**, 6400 (2023).

52 G. Sala, J. Meyer, A. Flechsig, L. Gabriel, P. Gambardella, Deterministic and stochastic aspects of current-induced magnetization reversal in perpendicular nanomagnets, Phys. Rev. B. **107**, 214447 (2023).

53 C. Zhang, S. Fukami, H. Sato, F. Matsukura, H. Ohno, Spin-orbit torque induced magnetization switching in nano-scale Ta/CoFeB/MgO, Appl. Phys. Lett. **107**, 012401 (2015).

54 S. Fukami, H. Ohno, Magnetization switching schemes for nanoscale three-terminal spintronics devices, Japanese Journal of Applied Physics. **56**, 0802A1 (2017).

55 K. Garello, I. M. Miron, C. O. Avci, F. Freimuth, Y. Mokrousov, S. Blügel, S. Auffret, O. Boulle, G. Gaudin, P. Gambardella, Symmetry and magnitude of spin–orbit torques in ferromagnetic heterostructures, Nat. Nanotechnol. **8**, 587 (2013).

56 D.-K. Lee, K.-J. Lee, Spin-orbit torque switching of perpendicular magnetization in ferromagnetic trilayers, Sci. Rep. **10**, 1772 (2020).

57 Z. Fu, A. K. Shukla, Z. Mu, K. Lu, Y. Dong, Z. Luo, X. Qiu, Z. Zhu, Y. Yang, Optimal Spin Polarization for Spin–Orbit-Torque Memory and Oscillator, IEEE Transactions on Electron Devices. **69**, 1642 (2022).

58 H. Akimoto, H. Kanai, Y. Uehara, T. Ishizuka, S. Kameyama, Analysis of thermal magnetic noise in spin-valve GMR heads by using micromagnetic simulation, J. Appl. Phys. **97**, 10N705 (2005).

# Figures

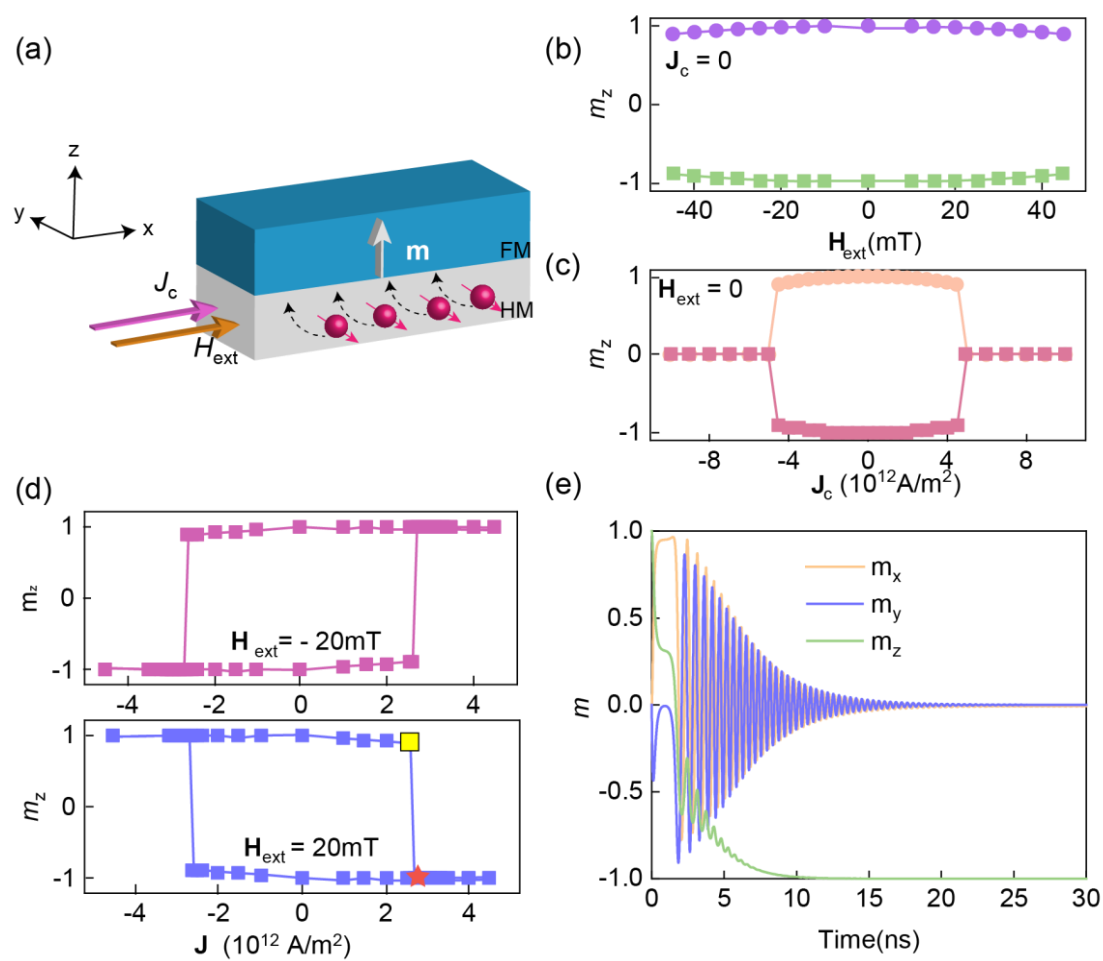

Fig. 1 (a) The illustration of device structure. (b) $m_z$ as a function of $\mathbf{H}_{ext}$ when $\mathbf{J}_c = 0$. The dots and squares represent the sample with the initial state $m_z = 1$ and $m_z = -1$, respectively. (c) $m_z$ as a function of $\mathbf{J}_c$ when $\mathbf{H}_{ext} = 0$. (d) $m_z$-$\mathbf{J}_c$ loop with negative (purple) and positive (blue) $\mathbf{H}_{ext}$. The yellow square and the star mark points before and after the switching. (e) Time evolution of $\mathbf{m}$ when $\mathbf{J}_c = 2.604 \times 10^{12}$ A/m², corresponding to the point marked by star.

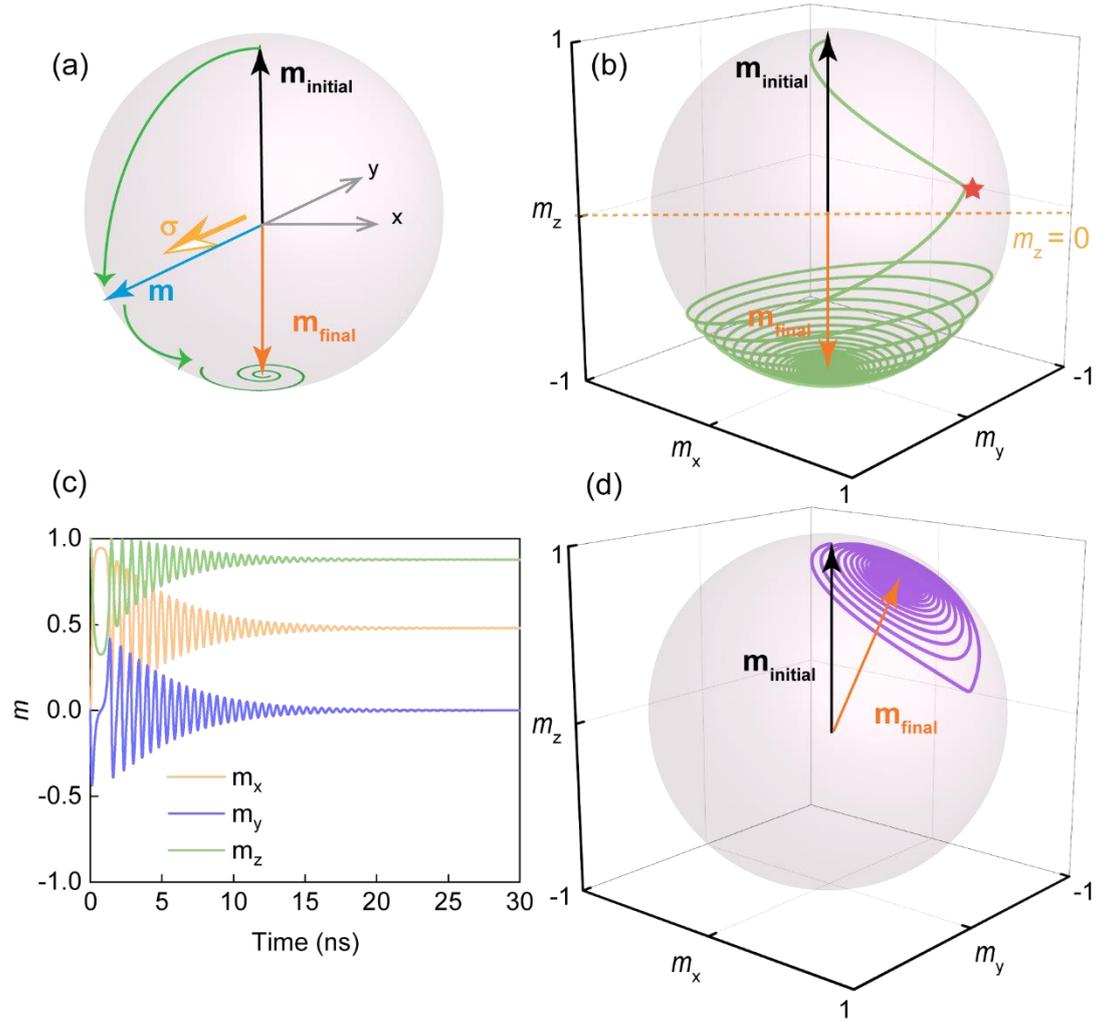

Fig. 2 Illustration of the switching process used in the previous model. (b) Switching trajectory obtained in the macrospin simulation under $\mathbf{J}_c = 2.604 \times 10^{12}$ A/m² and $\mathbf{H}_{ext} = 20$ mT. (c)-(d) Magnetization trajectory of the non-switched sample under $\mathbf{J}_c = 2.603 \times 10^{12}$ A/m² and $\mathbf{H}_{ext} = 20$ mT.

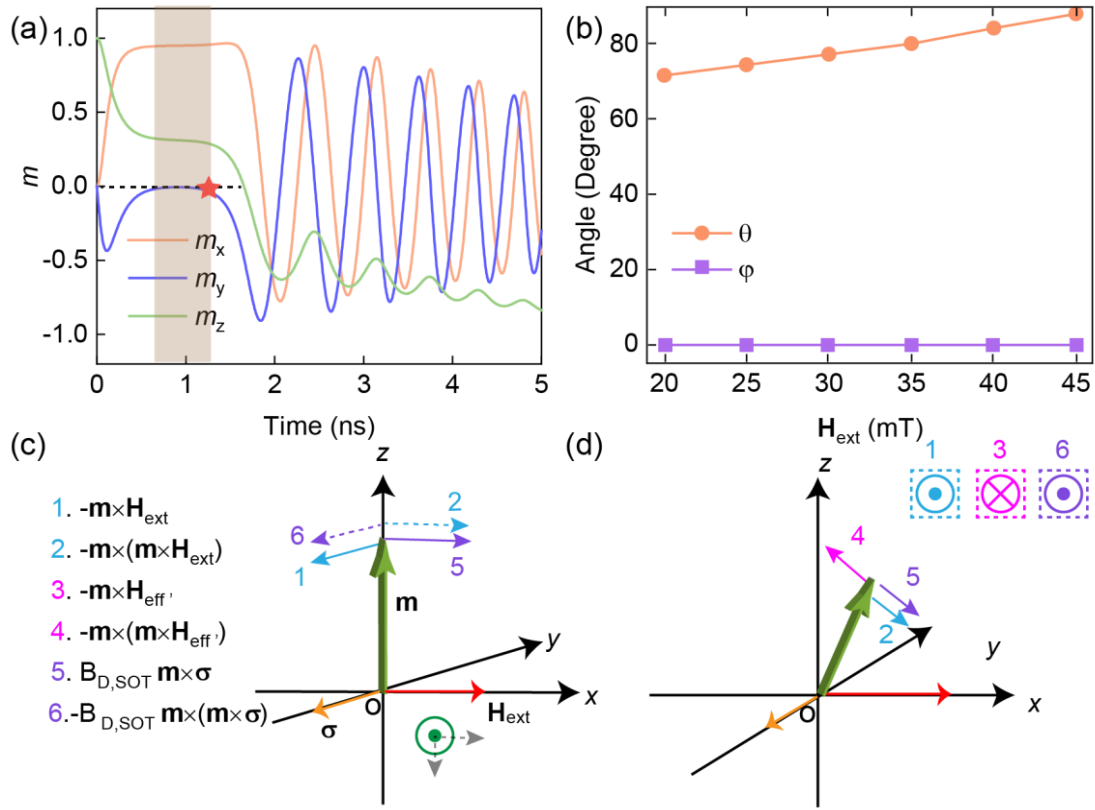

Fig. 3 (a) The zoomed magnetization trajectory, the same as Fig. 1(d). (b) $\theta$ and $\varphi$ at the chirality reversal point when different $\mathbf{H}_{ext}$ is applied. The relationship between $\theta$ and $H_{ext}$ is $\theta = 0.65 H_{ext} + 58.113$. The analysis of torques when the magnetization is (c) along +**z** direction and (d) in the **x-z** plane.

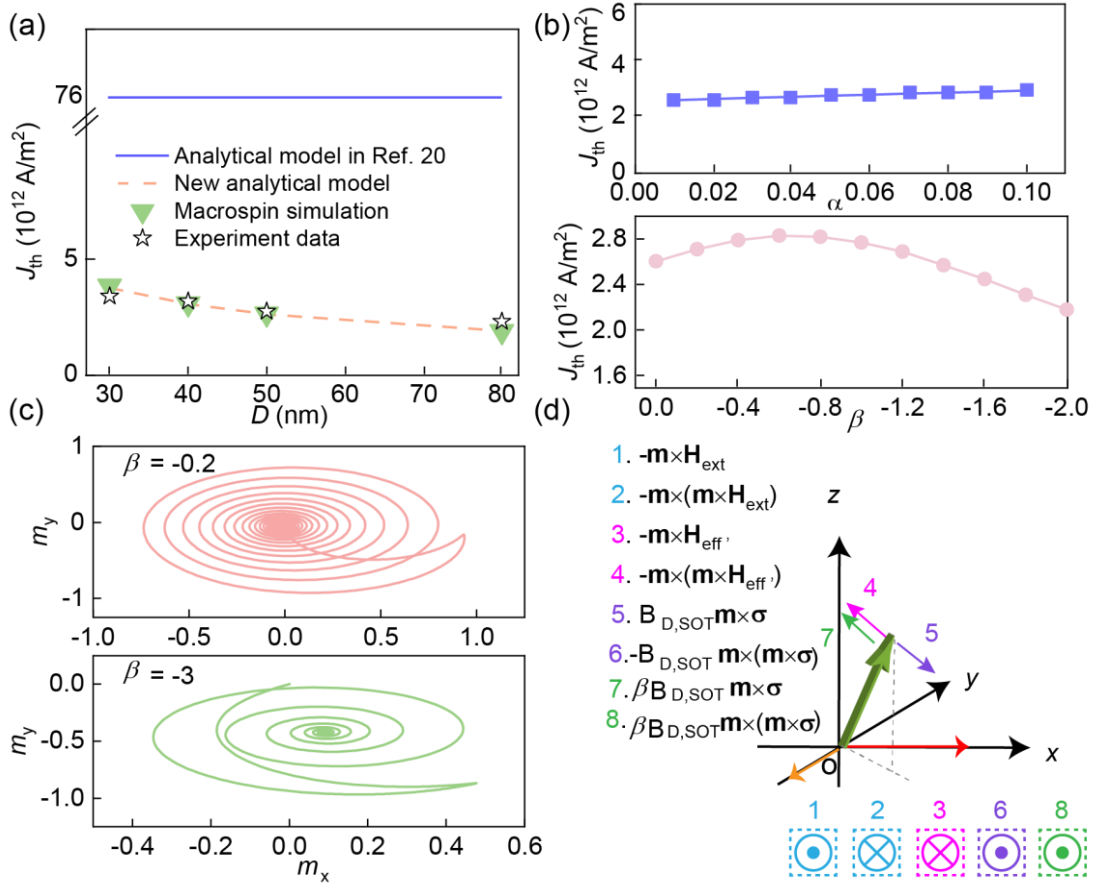

Fig. 4 $J_{th}$ as a function of (a) device size, (b) damping constant and field-like torque. (c) The trajectory of **m** projected into the **x-y** plane under different $\beta$. (d) The analysis of torques with the consideration of FLT.